\documentclass[aps,pra,showpacs,twocolumn,superscriptaddress]{revtex4}
\usepackage{graphicx}
\usepackage{amssymb}
\usepackage{epsfig}
\usepackage{graphicx}
\usepackage{amsmath}
\usepackage{array,color}

\usepackage{ulem}

\begin{document}
\title{Fulde-Ferrell-Larkin-Ovchinnikov state to Topological Superfluidity
Transition in Bilayer Spin-orbit Coupled Degenerate Fermi
Gas}
\author{Liang-Liang Wang}
\affiliation{Beijing National Laboratory for Condensed Matter
Physics, Institute of Physics, Chinese Academy of Sciences, Beijing
100190, China}
\author{Qing Sun}
\email{sunqing@cnu.edu.cn} \affiliation{Department of Physics,
Capital Normal University, Beijing 100048, China}
\author{W.-M. Liu}
\affiliation{Beijing National Laboratory for Condensed Matter
Physics, Institute of Physics, Chinese Academy of Sciences, Beijing
100190, China}
\author{G. Juzeli\={u}nas}
\email{gediminas.juzeliunas@tfai.vu.lt}\affiliation{Institute of
Theoretical Physics and Astronomy, Vilnius University, A.
Go\v{s}tauto 12, Vilnius 01108, Lithuania}
\author{An-Chun Ji}
\email{andrewjee@sina.com}\affiliation{Department of Physics,
Capital Normal University, Beijing 100048, China}
\date{{\small \today}}

\begin{abstract}
Recently a scheme has been proposed for generating the 2D Rashba-type spin-orbit coupling (SOC) for ultracold atomic bosons in a bilayer geometry [S.-W. Su et al, Phys. Rev. A \textbf{93}, 053630 (2016)]. Here we investigate the superfluidity properties of a degenerate Fermi gas affected by the SOC in such a bilayer system. We demonstrate that a Fulde-Ferrell-Larkin-Ovchinnikov (FFLO) state appears in
the regime of small to moderate atom-light coupling. In contrast to the ordinary SOC, the FFLO state emerges in the bilayer system without adding any external fields or spin polarization. As the atom-light
coupling increases, the system can transit from the FFLO state to a
topological superfluid state. These findings are also confirmed by the BdG simulations with a weak harmonic trap added.
\end{abstract}
\pacs{67.85.-d, 03.75.Mn, 05.30.Jp}







\maketitle
\section{Introduction}

The search for new exotic quantum states \cite{Hasan,Qi} are among
the most fundamental issues in current condensed matter physics.
These topics have drawn enormous interest for ultracold atomic gases
\cite{Lewenstein,Bloch,Dalibard,Lewenstein1,Goldman,Zoller} enabling
simulations of many condensed matter phenomena. With a recent experimental progress in synthetic SOC for degenerate atomic
gases \cite{Lin,ZhangJY,WangPJ,Cheuk,Qu,Huang,Meng,Wu,LiJ}, diverse new quantum phases due
to the SOC have been predicted \cite{Goldman,Zhai,Galitski}, such as
the stripe phase and vortex structure in the ground states
of atomic Bose-Einstein condensates (BECs)
\cite{WangCJ,Ho,XuZF1,Zhang,Li,Vyasanakere1,Sinha,Kawakami,Hu1,Wilson,Han,SunQ-2016}, as well as the Rashba pairing bound states (Rashbons) \cite{Jiang,Vyasanakere2}
and topological superfluidity
\cite{Hu2,Gong,Yu1,Gong1,Dong} in degenerate Fermi gases.

The synthetic SOC has been successfully implemented and explored by Raman coupling of a pair of atomic hyperfine ground states accompanied by a recoil  \cite{Lin,ZhangJY,WangPJ,Cheuk,Qu}.
This provides the SOC along the recoil direction representing the one-dimension (1D) SOC.
The realization of the synthetic SOC for ultracold atoms in two or more dimensions is
very desirable. The two
dimensional SOC of the Rashba type has a non-trivial
dispersion. The lower dispersion branch contains a highly degenerate ground state (the Rashba
ring). Additionally, there is a Dirac cone at an intersection point of two
dispersion branches, and a band gap can be opened by adding a Zeeman term. This is essential for the topological
superfluidity. Recently, a number of elaborate schemes has been
suggested to create an effective two- and three-dimensional (2D and
3D) SOC
\cite{Ruseckas,Stanescu1,Jacob,Stanescu2,Juzeliunas0,Juzeliunas1,Campbell,XuZF2,Anderson1,Anderson2,
XuZF3,LiuXJ,Sun,Su2}. Subsequently the 2D SOC has been
experimentally implemented \cite{Huang,Meng,Wu} by inducing the
Raman transitions between three atomic hyperfine ground states in a ring coupling scheme \cite{Huang,Meng,Campbell}, as well as by using another approach which relies on optical lattices \cite{Wu,LiuXJ}.
However in the experiments \cite{Huang,Meng} one of the three atomic states belongs to a higher hyperfine manifold leading to losses.

Recently Su et al \cite{Su2} put forward a scheme for generating the
effective Rashba SOC for a two-component atomic BEC confined in a bilayer geometry. The layer index provides an extra degree of freedom to form a basis of four combined spin-layer states composed of two spin and two layer states. The four spin-layer states are coupled in a cyclic manner by means of the spin flip Raman transitions \cite{Lin} and the laser-assisted interlay coupling \cite{Jaksch,Aidelsburger,Aidelsburger1,Miyake,Atala}, both processes being accompanied by recoil. This provides effectively a ring coupling scheme\cite{Su2} leading to the Rashba-type SOC. In contrast to the original ring coupling scheme \cite{Campbell} involving four atomic internal states, the bilayer setup makes use of only two atomic spin states\cite{Su2}, like the NIST scheme for the 1D SOC \cite{Lin}. Hence there is no need to employ spin states belonging to a higher
hyperfine manifold which suffers from a collisional population decay
undermining the effective SOC. Therefore, the bilayer scheme
offers a more feasible system to study the many-body physics due to
2D SOC.

Topological superfluidity has  attracted an enormous interest in SO coupled fermion gases \cite{Hu2,Gong,Yu1,Gong1,Dong,ZhangC,SatoM,QiXL,ZhengJH,XuY}. These works considered a pure 2D or 3D Rashba-type SOC for ultracold atoms with two internal (spin up and down) states. However, in a realistic atomic system the SOC is produced in an effective manner by laser dressing of a number of atomic internal states and restricting atomic motion to a manifold of a two fold degenerate dressed states.
As a result, the topological superfluidity is more involved, an issue which has been little investigated.

Here, we explore the topological superfluidity of a
Fermi gas affected by the SOC. The SOC is produced
using the bilayer scheme previously considered for bosonic atoms \cite{Su2,Xiong}. A characteristic feature of the bilayer system is
that the interaction takes place mostly between atoms belonging to the same layer \cite{Su2}. Therefore the atom-atom interactions differ considerably from the ones featured in the SOC scheme involving four cyclically coupled atomic internal states \cite{Campbell}. In the latter situation the atoms in all four internal states interact with approximately the same strength.

We find that the bilayer scheme provides an intriguing
phase transition of superfluidity. In the regimes of small and moderate
atom-light coupling the FFLO states
emerge. The FFLO states involve pairing with a finite centre-of-mass
momentum. These states have received a considerable attention in different physical contexts \cite{LiaoYA,HuH,IskinM,ChunleiQu,WeiZhang,Franca}. In general, the FFLO state arises in the spin-polarized
systems or can be induced by in-plane Zeeman fields in the SO coupled
Fermi gases \cite{ChunleiQu,WeiZhang}. Here we show that for the small to moderate atom-light coupling the FFLO states emerge intrinsically in the bilayer system without any external magnetic field or spin polarization.
As the atom-light coupling increases, the system can undergo a transition from the FFLO state to the topological superfluid (TS). The TSs have been pursued theoretically in model Hamiltonians with the 2D (3D) Rashba-type SOC \cite{Hu2,Gong,Yu1,Gong1,Dong,ZhangC,SatoM,QiXL,ZhengJH,XuY}. Here we
provide convincing evidences that TSs can also occur in the
experimentally feasible bilayer system.

\begin{figure}[t]
\centering
\includegraphics[width=0.45\textwidth]{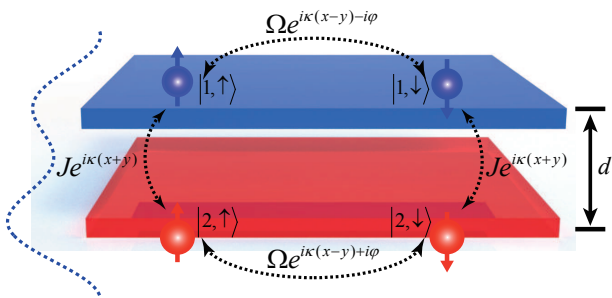}\\
\caption{(Color online) A schematic representation of a Fermi gas in a bilayer structure. An asymmetric double-well potential along the $z$ axis traps the atoms in two layers separated by a distance $d$. The combined spin-layer atomic  states  $\left|j=1,2,\gamma = \uparrow, \downarrow \right\rangle $ are cyclically coupled by the intralayer Raman transitions and interlayer laser-assisted tunneling characterized by the matrix elements  $\Omega e^{i\kappa(x-y)\pm i\varphi}$ and $J e^{i\kappa(x+y)}$, respectively.}\label{fig0}
\end{figure}

\section{The model and single particle spectrum}

We consider a two-component Fermi gas confined in
the bilayer geometry as illustrated in Fig. \ref{fig0}. The atoms are confined in a deep enough asymmetric double-well potential \cite{SebbyStrabley}, so their motion in the $z$-direction is restricted to the ground states of individual wells separated by a distance $d$. On the other hand, the laser-assisted tunneling can induce transitions between the two wells.

The system is
described by four combined spin-layer states $\left|j,\gamma\right\rangle =\left|j\right\rangle_{\rm
layer}\otimes\left|\gamma\right\rangle_{\rm spin}$ which serve as the
states required for the ring coupling scheme \cite{Campbell}. Here $j=1,2$ signifies
the $j$-th layer, and $\gamma=\uparrow,\downarrow$
denotes an internal atomic state. For example, the states with $\gamma=\uparrow,\downarrow$ can be the $m_F=9/2$ and $m_F=7/2$ magnetic sublevels of the $F=9/2$ hyperfine ground state
manifold of fermion $^{40}\mathrm{K}$ atoms, like in the experiment \cite{WangPJ} on the 1D SOC.

The four atomic states $\left|j,\gamma\right\rangle$ are coupled in a cyclic manner  via the spin-flip Raman transitions and the interlayer laser-assisted tunneling characterized by the matrix elements $\Omega \left( \mathbf{r} \right) =\Omega e^{i\kappa(x-y)\pm i\varphi}$ and $J\left( \mathbf{r} \right) = J e^{i\kappa(x+y)}$, respectively. As discussed in ref. \cite{Su2}, such Raman transitions and interlayer tunneling can be induced using three properly chosen laser beams.  The Raman coupling provides a recoil in the $x+y$ direction, whereas the interlayer tunneling is accompanied by a recoil in the $x-y$ direction. Here $\kappa\sqrt{2}$ is a magnitude of the in-plane momentum transferred by the lasers, and $2\varphi=k_{z}d$ is a phase difference of the Raman coupling between the two layers due to an out-of-plane recoil momentum $k_{z}$\cite{Su2}.

The recoil momentum $k_{z}$  can also influence the atomic center of mass motion in the z direction. However the latter effect is not important if the atoms are confined strongly by the potential wells in the z direction. This is justifiable if the depth of each well comprising the asymmetric double well potential exceeds considerably the recoil energy. Consequently  in each layer the atomic ground state is localized in the z direction over distances much smaller than the wave-length $\lambda_{z}=2\pi / k_{z}$ corresponding to the atomic recoil accompanying the spin flip transitions. In that case, the atoms remain in the ground states of individual potential wells after spin-flip transitions.

\subsection{Hamiltonian and system}
Performing a gauge transformation eliminating the position-dependence of the atom-light coupling matrix elements $\Omega \left( \mathbf{r} \right)$ and $J\left( \mathbf{r} \right) $, the bilayer system is described by the following  Hamiltonian \footnote{For more details see Appendix \ref{AppendixA}, as well as Appendix A of ref. \cite{Su2}.}:
\begin{equation}
\hat{H}=H_{\rm kin}+H_{\rm Laser}+H_{\rm SOC}+H_{\rm int}\,,
\label{H-sec-quant}
\end{equation}
where
\begin{eqnarray}
H_{\rm kin}&=&\int
d^2\boldsymbol{\mathbf{r}}\Sigma_{j,\gamma}\hat{\psi}^{\dagger}_{j\gamma} \left[\frac{\hbar^{2}k^{2}}{2m}-\mu\right]\hat{\psi}_{j\gamma}\,,\\
H_{\rm Laser}&=&\int d^2\boldsymbol{\mathbf{r}}
\Omega\left[e^{i\varphi}\hat{\psi}^{\dagger}_{1\uparrow}
\hat{\psi}_{1\downarrow}
+e^{-i\varphi}\hat{\psi}^{\dagger}_{2\uparrow}
\hat{\psi}_{2\downarrow}+{\rm H.c.}\right]\notag\nonumber\\
&+&\int d^2\boldsymbol{\mathbf{r}}\Sigma_\gamma J
\hat{\psi}^{\dagger}_{2\gamma}
\hat{\psi}_{1\gamma}+{\rm H.c.}\,, \label{H_Laser} \\
 H_{\rm SOC}&=&\int d^2\boldsymbol{\mathbf{r}}
\frac{\hbar^{2}\kappa}{m}\left[\hat{\psi}^{\dagger}_{2\uparrow}
k_x\hat{\psi}_{2\uparrow}
-\hat{\psi}^{\dagger}_{1\downarrow}k_x\hat{\psi}_{1\downarrow}
\right.\notag\nonumber\\
 &&\>\>\>\>\>\>\>\>\>\>\>\>\>\>\>\>\>\>+\left.\hat{\psi}^{\dagger}_{2\downarrow}
k_y\hat{\psi}_{2\downarrow}
-\hat{\psi}^{\dagger}_{1\uparrow}k_y\hat{\psi}_{1\uparrow}
\right]\,,\nonumber\\
H_{\rm int}&=&-g\int
d^2\boldsymbol{\mathbf{r}}\sum_{j}\hat{\psi}^{\dagger}_{j\uparrow}
\hat{\psi}^{\dagger}_{j\downarrow}\hat{\psi}_{j\downarrow}
\hat{\psi}_{j\uparrow} \,.\label{Hamiltonian1}
\end{eqnarray}
where $\hat{\psi}_{j\gamma}\equiv \hat{\psi}_{j\gamma}({\mathbf r})$
is a fermion field operator for annihilation of an atom positioned at ${\mathbf r}$
in a layer $j=1,2$ with a spin state $\gamma =\downarrow\,,\uparrow$.
The first term  $H_{\rm kin}$ represents the Hamiltonian for an unperturbed atomic
motion within the layers,  $\mu$ being the chemical
potential. The second term $H_{\rm Laser}$ accommodates the
spin-flip intralayer Raman transitions characterized by the Rabi
frequency $\Omega$, as well as the laser-assisted interlayer tunneling described by the
strength $J$. The third term $ H_{\rm SOC}$ represents the SOC
due to the recoil momentum $\kappa$ in the $xy$ plane induced by the
interlayer tunneling and Raman transitions \cite{Su2}. The latter SOC term was not included in the previous analysis of the
superfluidity for the fermions in a bilayer geometry \cite{ZhengJH}.

\begin{figure}[t]
\centering
\includegraphics[width=0.45\textwidth]{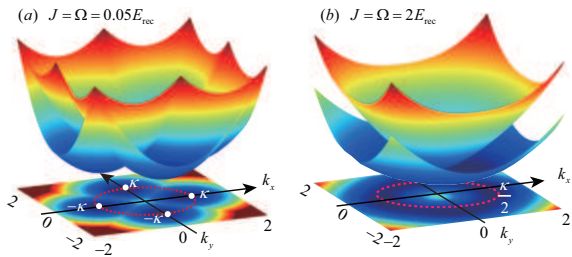}\\
\caption{(Color online). Single-particle dispersion of the lowest two
branches for various coupling strengths and $\varphi=\pi/2$. (a) In
a weak coupling regime $\Omega=J\ll E_{\rm rec}$, the dispersion is
a superimposition of four distinct paraboloids nearly centered at
$(\pm \kappa,\pm \kappa)$. (b) In the strong coupling regime
$\Omega=J\gg E_{\rm rec}$, the Rashba-ring minimum (red dashed line
) with a radius $\kappa/2$ emerges.}\label{fig1}
\end{figure}

As one can see in Fig. \ref{fig0} and Eq. (\ref{H_Laser}), the amplitude
of the Raman coupling $\Omega e^{\pm i \varphi}$ contains a relative phase $2\varphi=k_{z}d$ between
the upper and lower layers. The phase $2\varphi$ can be changed by either varying the double-well separation $d$ or the out-of-plane Raman recoil momentum $k_{z}$.
For  $\varphi=\pi/2$ a Dirac cone appears in the single particle spectrum of the ring
coupling scheme \cite{Campbell,Su2}. A gap is opened in the Dirac spectrum for $\varphi\ne\pi/2$.
This is important for formation of the topological superfluidity.

Finally, $H_{\rm int}$ describes the on-site
attractive interaction ($g>0$)  between atoms situated in the same layer.  Here the
bare s-wave interaction  $g$ is related to the binding energy
$E_{b}$ of the two-body bound state \cite{Shieh} via
\begin{equation}
1/g=\sum_{\mathbf{k}}1/(2\varepsilon_{\mathbf{k}}-E_{b})\,,
\label{1/g}
\end{equation}
where $\varepsilon_{\mathbf{k}}=\hbar^2\mathbf{k}^2/(2m)$ is the
kinetic energy. In experiments, the binding energy $E_{b}$
can be tuned via the Feshbach resonance technique.
\begin{figure}[t]
\centering
\includegraphics[width=0.42\textwidth]{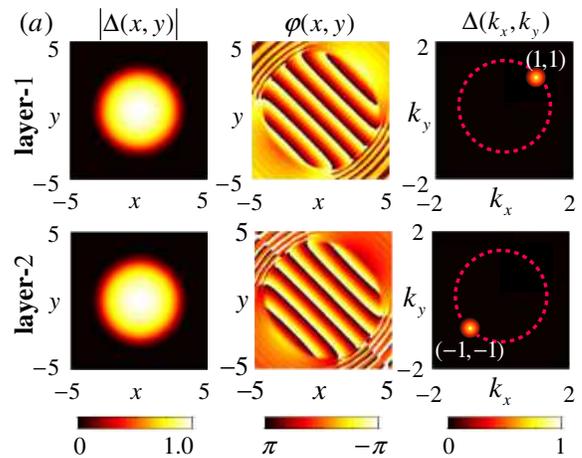}\\
\caption{(Color online) (a) The real-space density profile (left
panel) and phase configurations (middle panel) of the order
parameters $\Delta_{1,2}$ for $\Omega=J=0.05E_{\mathrm{rec}}$ and
$\varphi=\pi/2$.  Here we have taken $N_{\mathrm{atom}}=100$ and
$E_{b}=2E_{\mathrm{rec}}$. The right panel shows the corresponding
momentum-space distribution. (b) An illustration of the FFLO-type of the
Cooper pairing mechanism. The red and blue solid arrows represent
the Cooper pairing momenta of atoms in the first and second layers,
with $\mathbf{Q}_{1,2}$ denoting the total paring momentum.
}\label{fig2}
\end{figure}
\begin{figure*}
\includegraphics[width=0.95\textwidth]{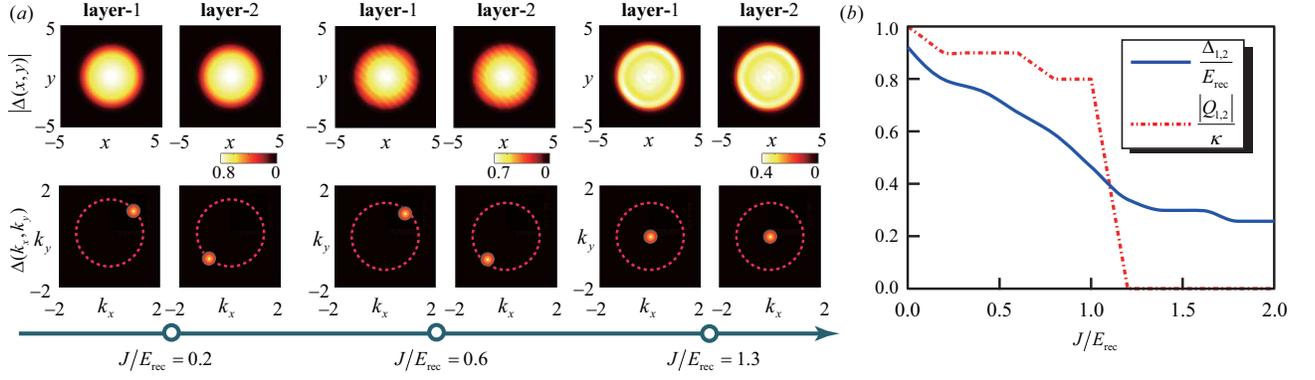}
\caption{(Color online). (a) Profiles of the order parameter
$\Delta(\mathbf{r})$ (up) and the corresponding momentum distributions
(down) in the first and second layers for the $\Omega=J=0.2, 0.6,
1.3E_{\mathrm{rec}}$. Other parameters are $E_{b}=2E_{\mathrm{rec}}$,
$\varphi=\pi/2$, and $N_{\mathrm{atom}}=100$. (b) Evolution of the
order parameter $\Delta_{1,2}$ (maximum of the order parameter in the whole region) (blue solid line) and the magnitude of the FFLO vector $|\mathbf{Q}_{1,2}|$
(red dashed-dotted line) are plotted as a function of the interlayer tunneling strength
$J=\Omega$.}\label{fig3}
\end{figure*}

\subsection{Effective single-particle Hamiltonian.}

We shall focus on a
situation of a symmetric coupling: $\Omega=J$.  Figure \ref{fig1}
plots the single-particle dispersion. If $\Omega=J\ll E_{\rm rec}$, the
minima of the single-particle dispersive paraboloids appear at
$(\pm \kappa,\pm \kappa)$, where $E_{\rm
rec}=\hbar^2\kappa^2/2m$ is a characteristic energy of the in-plane recoil. The dispersion
 is then built of four distinct superimposed
paraboloids, each corresponding to individual spin-layer states
$\left|j,\gamma\right\rangle$. As the coupling
increases (but $\Omega=J\leq E_{\rm rec}$), the four paraboloids
gradually coalesce. Yet the dispersion still exhibits four
distinguishable minima in the lowest branch. The locations of the
four minima would gradually move towards to center $k=0$ with increasing the coupling. Finally
in the strong coupling regime, $\Omega=J\gg E_{\rm rec}$, the mixing
between the spin states results in the emergence of a cylindrically
symmetric Rashba-ring minimum of a radius $\kappa/2$ \cite{Su2,Campbell}. In this
case, one can project the Hamiltonian onto the lowest two energy
states leading to the usual single particle Hamiltonian of the Rashba-type (see Appendix \ref{AppendixA})
\begin{equation}
H_{\rm eff}=
\begin{pmatrix}
\mathbf{k}^2/2m-\mu'-h_{z} & \alpha(k_{y}-ik_{x})\\
\alpha(k_{y}+ik_{x}) & \mathbf{k}^2/2m-\mu'+h_{z}\\
\end{pmatrix}\,,
\label{H_eff}
\end{equation}
where $
\mu'=\mu-\Delta \mu $ is an effective chemical potential, $\alpha=\kappa/2m$ and $\Delta \mu=E_{\mathrm{rec}}-\Omega \sqrt{2} \cos{\left( \delta \varphi /2\right)}$.
Here
\begin{equation}
h_{z}
=\Omega \sqrt{2} \sin{\left( \delta \varphi /2\right)}\,, \quad \mathrm{with} \quad \delta \varphi =  \varphi -\pi/2,
\label{H_z-1}
\end{equation}
is an
effective Zeeman field which is controled by tuning the relative phase $\varphi$ for the Raman coupling. For $\varphi=\pi/2$ we have $h_{z}=0$, leading to the usual Dirac cone at $k=0$. The Dirac cone is opened if the
phase $\varphi$ deviates from $\pi/2$.

\section{Analysis of the superfuildity}

\subsection{Method}

By introducing the superfluid order parameters
$\Delta_j(\textbf{r})=g\left<\psi_{j,\downarrow}(\mathbf{r})
\psi_{j,\uparrow}(\mathbf{r})\right>$ with $\mathbf{r}=(x,y)$, the Hamiltonian
(\ref{Hamiltonian1}) can be diagonalized via a
Bogoliubov--Valatin transformation \cite{Valation}. In doing so we have additionally included a weak
harmonic trapping potential $V(r)=m\omega^2r^2/2$. The resultant Bogoliubov--de Gennes
(BdG) equation
\begin{equation}
H_{\rm BdG}(\mathbf{r})\phi_{\eta}=\varepsilon_{\eta}\phi_{\eta}
\label{H_BdG-Eq}
\end{equation}
is described by an $8\times 8$ matrix Hamiltonian
\begin{equation}
H_{\mathrm{BdG}}(\mathbf{r})=\left( \begin{matrix}H_{1}(\mathbf{r}) & \mathcal{J}\\
\mathcal{J}^{\dagger} & H_{2}(\mathbf{r})\end{matrix}\right)\,,
\label{H_BdG}
\end{equation}
where a diagonal $4\times4$ matrix $\mathcal{J}=\mathrm{diag}(J,J,-J,-J)$ describes the inter-layer coupling, and $H_{1,2}(\mathbf{r})$ is a $4\times4$ matrix Hamiltonian for an uncoupled layer $j=1,2$:
\begin{equation}
\begin{split}
H_{1}(\mathbf{r}) &=
\begin{pmatrix}
\epsilon_{1\uparrow}(\mathbf{r}) & \Omega e^{-i\varphi} & 0 & \Delta_{1}(\mathbf{r})\\
\Omega e^{i\varphi} & \epsilon_{1\downarrow}(\mathbf{r}) & -\Delta_{1}(\mathbf{r}) & 0\\
0 & -\Delta_{1}^{\ast}(\mathbf{r}) & -\epsilon_{1\uparrow}^{\ast}(\mathbf{r}) & -\Omega e^{-i\varphi}\\
\Delta_{1}^{\ast}(\mathbf{r}) & 0 & -\Omega e^{i\varphi} & -\epsilon_{1\downarrow}^{\ast}(\mathbf{r})\\
\end{pmatrix}\,,\\
\end{split}
\end{equation}
and
\begin{equation}
\begin{split}
H_{2}(\mathbf{r}) &=
\begin{pmatrix}
\epsilon_{2\uparrow}(\mathbf{r}) & \Omega e^{i\varphi} & 0 & \Delta_{2}(\mathbf{r})\\
\Omega e^{-i\varphi} & \epsilon_{2\downarrow}(\mathbf{r}) & -\Delta_{2}(\mathbf{r}) & 0\\
0 & -\Delta_{2}^{\ast}(\mathbf{r}) & -\epsilon_{2\uparrow}^{\ast}(\mathbf{r}) & -\Omega e^{i\varphi}\\
\Delta_{2}^{\ast}(\mathbf{r}) & 0 & -\Omega e^{-i\varphi} & -\epsilon_{2\downarrow}^{\ast}(\mathbf{r})\\
\end{pmatrix}\,,\\
\end{split}
\label{BdGequation}
\end{equation}
with
\begin{equation}
\left\{
\begin{aligned}
\epsilon_{1\uparrow}(\mathbf{r})&=-\hbar^{2}\nabla^2/(2m)+i\hbar^{2}\kappa\partial_{y}/m+V(\mathbf{r})-\mu\\
\epsilon_{1\downarrow}(\mathbf{r})&=-\hbar^{2}\nabla^2/(2m)+i\hbar^{2}\kappa\partial_{x}/m+V(\mathbf{r})-\mu\\
\epsilon_{2\uparrow}(\mathbf{r})&=-\hbar^{2}\nabla^2/(2m)-i\hbar^{2}\kappa\partial_{x}/m+V(\mathbf{r})-\mu\\
\epsilon_{2\downarrow}(\mathbf{r})&=-\hbar^{2}\nabla^2/(2m)-i\hbar^{2}\kappa\partial_{y}/m+V(\mathbf{r})-\mu\\
\end{aligned}
\right.\,.
\end{equation}
The Nambu basis is chosen as $\phi_{\eta}=[u_{1\uparrow,\eta},u_{1\downarrow,\eta},
v_{1\uparrow,\eta},v_{1\downarrow,\eta},u_{2\uparrow,\eta},u_{2\downarrow,\eta},
v_{2\uparrow,\eta},v_{2\downarrow,\eta}]^{T}$, and $\varepsilon_{\eta}$ is the corresponding energy of
Bogoliubov quasiparticles labeled by an index ``$\eta$''.
The order parameter $\Delta_{1,2}(\mathbf{r})$ is to be determined self-consistently by
\begin{equation}
\Delta_{j}(\mathbf{r})=g\sum_{\eta}[u_{j\uparrow,\eta}
v^{\ast}_{j\downarrow,\eta}f(-\varepsilon_{\eta})\nonumber+u_{j\downarrow,\eta}v^{\ast}_{j\uparrow,\eta}f(\varepsilon_{\eta})]\,,
\end{equation}
where $f(E)=1/[e^{E/k_{B}T}+1]$ is the Fermi-Dirac distribution function
at a temperature $T$. The chemical potential $\mu$ is determined using
the number equation $N=\int d\mathbf{r} n(\mathbf{r})$, where the
total atomic density is given by
\begin{equation}
n(\mathbf{r})=\sum_{j\gamma,\eta}[|u_{j\gamma,\eta}(\mathbf{r})|^{2}
f(\varepsilon_{\eta})+|v_{j\gamma,\eta}(\mathbf{r})|^{2}f(-\varepsilon_{\eta})]\,.
\end{equation}
We have solved the BdG equation self-consistently by using the basis expansion method \cite{XuY}. In the numerical simulations we take a large energy cutoff
$\varepsilon_{c}=6E_{\mathrm{rec}}$ to ensure the accuracy of the calculation, where
$E_{\mathrm{rec}}=10\hbar\omega$ assures the trap oscillation
frequency is much smaller than the recoil frequency.
Throughout this work we focus on the case of zero temperature. The temperature of the gas $T$ is about $0.2\sim0.3 T_{\mathrm{F}}$ for a total number of atoms $N_{\mathrm{atom}}=100$ used in the calculations if one takes the Fermi temperature $T_{\mathrm{F}}$ to be around $300\mathrm{nK}$ \cite{Onofrio}. Such a temperature $T$ is sufficiently low, so the results of the calculation are almost unaffected by taking the zero temperature limit.
\subsection{Results}

We start with a weak coupling limit $\Omega=J\ll
E_{\rm rec}$. Figure \ref{fig2} plots the corresponding density
profiles and phase configurations for the order parameter
$\Delta_{1,2}(\mathbf{r})$. The Fermions are assumed to
populate only the lowest branch, so the Fermi surface forms four Fermi
pockets centered at $(\pm \kappa,\pm \kappa)$. The first two points
$(\kappa,0)$ and $(0,\kappa)$
correspond to the spin up and
down states for the first layer, whereas the remaining two points $(-\kappa,0)$ and $(0,-\kappa)$
correspond to the spin up and down states for the second layer. Note
that atoms prefer to pair in the same layer, as only the atoms situated in the same layer interact.
We have found  that the order
parameters $\Delta_{1,2}(\mathbf{r})=\Delta_0 e^{i
\mathbf{Q}_{1,2}\cdot\mathbf{r}} $ exhibit an oscillating structure along the diagonal
directions $\mathbf{Q}_1=\kappa(\mathbf{e}_{x}+\mathbf{e}_{y})$
and $\mathbf{Q}_2=-\mathbf{Q}_1$, as shown in Fig. \ref{fig2}(a).
This is a key  feature of the so-called FFLO phase. In Fig.
\ref{fig2} (b), we illustrate the underlying pairing mechanism of the FFLO
state.

If we choose the first layer, the wave vectors for the two pockets $(\kappa,0)$ and $(0,\kappa)$ can be written
as $\mathbf{k}_{1\uparrow}=\kappa\mathbf{e}_{y}+
\mathbf{K}_{1\uparrow}$ and
$\mathbf{k}_{1\downarrow}=\kappa\mathbf{e}_{x}+
\mathbf{K}_{1\downarrow}$. Here
$\mathbf{K}_{1\uparrow(\downarrow)}$ denotes the atomic momentum
calculated with respect to the center of each Fermi pocket. When $\mathbf{K}_{1\uparrow}=-\mathbf{K}_{1\downarrow}$, fermions can pair together in different pockets with opposite momenta.
In this case the total momentum $\mathbf{k}_{1\uparrow}+\mathbf{k}_{1\downarrow}$
of the atomic pair in the first layer
reads
$\mathbf{Q}_1=\kappa(\mathbf{e}_{x}+\mathbf{e}_{y})$. In a similar manner,
the pairing center-of-mass of
momentum is
$\mathbf{Q}_2=-\kappa(\mathbf{e}_{x}+\mathbf{e}_{y})$ for the second layer, as one can see in Fig. \ref{fig2}(b).

As the coupling $\Omega=J$ increases, the four paraboloids are mixed
by the intralayer spin flip atomic transitions and interlayer tunneling. In that case
the four Fermi surface pockets still remain, but the central locations of
the pockets shrink towards the momentum center $k=0$. In this case, the FFLO state sustains, but with
the FFLO pairing amplitude $\Delta_0$ being reduced. Fig. \ref{fig3}
depicts the evolution of the order parameters $\Delta_{1,2}$ and
corresponding momentum dispersions $Q_{1,2}$ for various coupling
strengths. We see that, the pairing momentum decreases with
increasing the coupling. Around $J\gtrsim E_{\mathrm{rec}}$, the four
paraboloids begin to merge together, the atoms tend to pair with
zero relative momentum and the system enter a normal superfluid
state.
\begin{figure}[t]
\centering
\includegraphics[width= 0.5\textwidth]{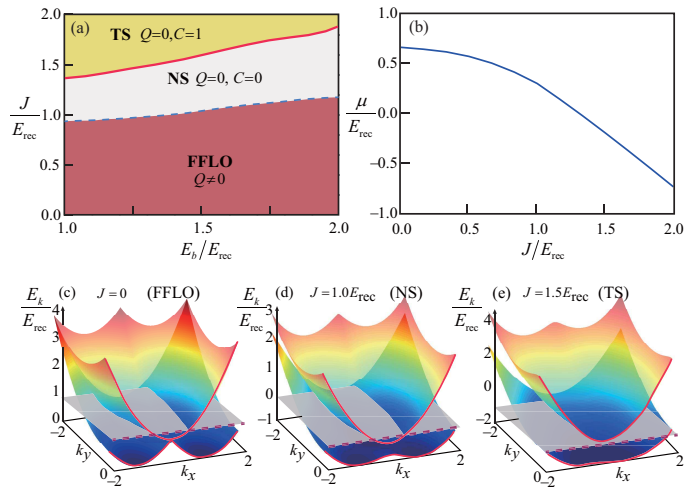}\\
\caption{(Color online) Phases of the bilayer system for $\Omega=J$ and  $\varphi=0.6\pi$.  (a) Phase diagram in the
$(E_{b},J)$ plane for $N_{\mathrm{atom}}=100$. Three phases are formed: a FFLO superfluid represented by a dark red region, a normal superfluid (NS, white region) and a topological
superfluid (TS, yellow region). The latter phase is characterized by a non-zero Chern number $C$ and a zero pairing momentum $Q$. (b) The chemical potential $\mu$ as a function of the interlayer tunneling $J$ at the binding energy $E_{b}=1.0E_{\mathrm{rec}}$. (c)-(e) Two lowest branches of the single-particle dispersion for different values of $J$ corresponding to different phases. Here the chemical potential is represented by a gray plane.
The dispersion branches become lower with an increase of J, leading to a decrease of the chemical potential, as one can see in (b)-(e).
The NS phase transforms to the TS phase when the chemical potential enters the energy gap, as one can see comparing (d) and (e).
In (c)-(e) the momentum is measured in the units of the recoil momentum $\kappa$.
}\label{fig4}
\end{figure}

Let us now investigate a feasibility of the topological superfluidity in the bilayer system for $\Omega=J$. For this purpose, we allow the relative phase of the Raman coupling $\varphi$ to
deviate from $\pi/2$, so $\delta\varphi= \varphi-\pi/2\ne 0$. Consequently the energy gap $E({\delta\varphi})=2h_z=2\sqrt{2}\Omega\sin(\delta\varphi /2)$
 appears in the single-particle spectrum at the Dirac point ${\bf k}=0$ in the strong coupling (Rashba) regime.
The phase diagram is illustrated in Fig. \ref{fig4}. As anticipated, at the strong coupling one can find a TS phase which is characterized by a non-zero Chern number $C=1$ and a zero pairing momentum $Q=0$. Here the Chern number is calculated by a self-consistent solution of the BdG equation (\ref{H_BdG-Eq}) at the center of the trap where the changes in the trapping potential are minimum (see the Appendix \ref{AppendixB}).
As shown in Fig. \ref{fig4}, there is a phase transition from normal to topological superfluids when the chemical potential enters the energy gap.
Moreover, the FFLO and NS states have also been identified by obseving the states with $Q\neq0$ and $Q=0, C=0$, respectively. For the moderate coupling, $\Omega=J<E_{\mathrm{rec}}$, the bilayer system stays in the FFLO state. Similar to the $\varphi=\pi/2$ case discussed above, such a FFLO state would undergo a transition to the NS when the coupling strength is increased to $J_{c1}\gtrsim E_{\mathrm{rec}}$. In the strong coupling regime, $\Omega=J\gtrsim 1.5E_{\mathrm{rec}}$, corresponding to the limit of the effective Rashba-type SOC,  the system can be in a TS state if the chemical potential is situated within the energy gap $E(\delta\phi)$. Hence there exists another critical coupling strength $J_{c2}$ for the transition between between the NS and TS.
In Fig.~\ref{fig4} the dashed blue and red solid lines indicate the phase transitions FFLO$\rightarrow$ NS and NS$\rightarrow$TS, respectively.
It is noteworthy that the critical coupling $J_{c2}$ for the transition between the NS and TS states increases with the binding energy $E_{b}$ because of the increase of the chemical potential.
\begin{figure}[t]
\includegraphics[width= 0.5\textwidth]{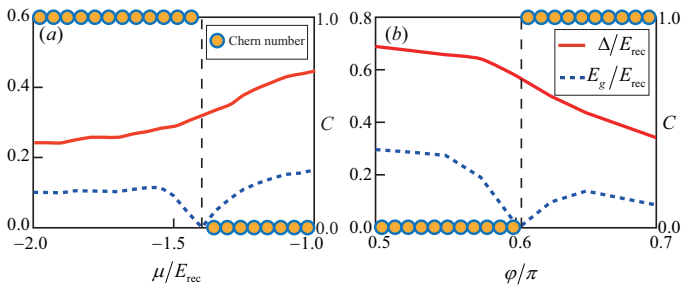}
\caption{(Color online)  Evolution of parameters of the system across the
topological phase boundary. (a) The behaviour of the minimum excitation
gap $E_{\mathrm{g}}$ (blue dashed line) and the order parameter $\Delta_{1,2}=\Delta$
(red solid line) with increasing the chemical potential for $E_{b}=3E_{\mathrm{rec}}$, $J\equiv \Omega=2E_{\mathrm{rec}}$, and
$\varphi=3/4\pi$. (b) The minimum excitation gap and
pairing order parameter against the tunneling phase $\varphi$ for $E_{b}=3E_{\mathrm{rec}}$, $J\equiv \Omega=2E_{\mathrm{rec}}$,
and $\mu=-0.5E_{\mathrm{rec}}$. The yellow filled circles indicate the integer
Chern number $C$ which is a topological
invariant.}\label{fig5}
\end{figure}

To explore in detail the transition from non-topological (NS) to
topological (TS) phases, we have observed the closing and reopening of the excitation
gap $E_{\mathrm{g}}$, which is necessary to change the topology of the Fermi
surface. In Fig. \ref{fig5}, we present a behavior of the order
parameter $\Delta_{1,2}=\Delta$, the bulk  quasi-particle gap $E_{\mathrm{g}}$ and the Chern
number $\mathcal{C}$ with increasing the chemical potential $\mu$
and the interlayer phase $\varphi$. One can see that across the
critical point where the Chern number changes abruptly, the
excitation gap $E_{\mathrm{g}}$ vanishes, indicating the topological phase
transition. The order parameter $\Delta$ increases with increasing
the chemical potential $\mu$, and decreases with an increase
of the interlayer phase $\varphi$ with respect to $\pi/2$.

\begin{figure}[t]
\includegraphics[width= 0.42\textwidth]{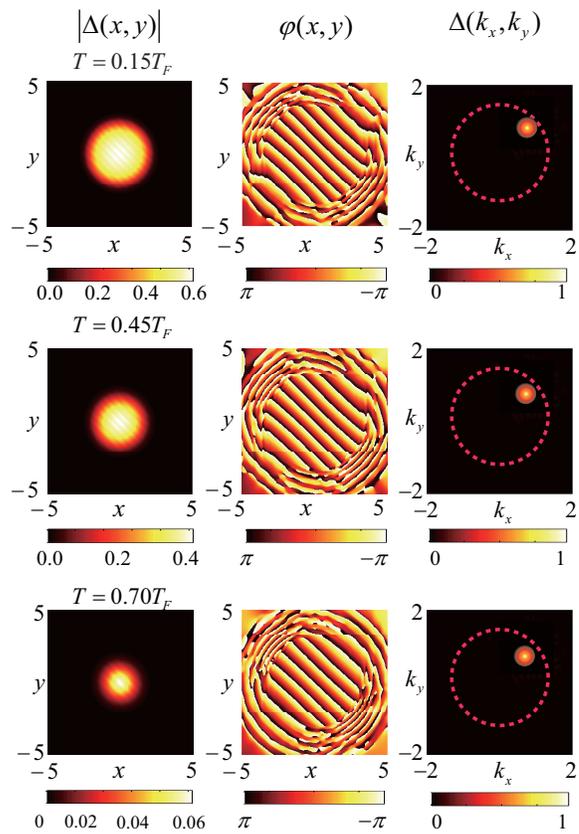}
\caption{(Color online) The density profiles, the phase configurations, and corresponding momentum-space distributions of the order parameter of the first layer at several temperatures: $T=0.15T_{F}$ (upper panel), $T=0.45T_{F}$ (middle panel), and $T=0.7T_{F}$ (bottom panel). Other parameters are $\Omega=J=0.4E_{\mathrm{rec}}$, $E_{b}=1E_{\mathrm{rec}}$, $\varphi=0.6\pi$ and $N_{\mathrm{atom}}=100$. The characteristic temperature scale $T_{F}=E_{F}/k_{B}$ is given by the Fermi energy.
The momentum is measured in the units of the recoil momentum $\kappa$.
}\label{fig6}
\end{figure}

Finally, we show how the FFLO state evolves with increasing the temperature in this bilayer system.  Fig.\ref{fig6} gives BdG results for the density profiles, the phase configurations, as well as the momentum-space distributions of the order parameter for the first layer at different temperatures. (The order parameter looks similar in the second layer, so we have not displayed such plots.) With increasing the temperature the order parameter is destroyed gradually, and a large FFLO momentum $\approx(0.9\kappa,0.9\kappa)$ is nearly independent of the temperature. In this case, it is much more difficult to disturb the FFLO state by thermal fluctuations compared to the TS phase. A further increase of the temperature will destroy the superfluid state eventually at around $T\approx 0.8T_{F}$.

\section{Concluding remarks}

We have investigated the superfluidity properties of
a bilayer spin-orbit coupled degenerate Fermi gas. The analysis has
elucidated a diverse phase diagram of the bilayer superfluidity in a
wide range of magnitudes of the atom-light coupling and atom-atom
interaction. For the small to moderate atom-light coupling, the FFLO states occur.
In the stronger coupling regime, the system undergoes a transition from  the NS to TS phases. These effects can be experimentally detected for atomic fermions in a
bilayer system. As discussed previously in the context bosons \cite{Su2},  the bilayer scheme can be readily realized using three laser beams to induce the interlayer tunneling and the intralayer spin flip transitions. In a similar manner, the bilayer scheme can be implemented for fermion atoms used in the previous experiments on the 1D SOC, such as $^{40}\mathrm{K}$ \cite{WangPJ}.

\begin{acknowledgments}
This work is supported by NSFC under grant Nos. 11474205, 11404225, 11504037, NKBRSFC under
grants Nos. 2012CB821305. G. J. acknowledges a support by the Lithuanian
Research Council (Grant No. MIP- 086/2015).
\end{acknowledgments}

\appendix
\begin{section}{single-particle Hamiltonian} \label{AppendixA}
A single particle
part of the bilayer Hamiltonian (\ref{H-sec-quant}) reads
in the matrix representation
\begin{equation}
H_0=
\begin{pmatrix}
\epsilon_{1\uparrow}(\mathbf{k}) & \Omega e^{-i\varphi} & J & 0 \\
\Omega e^{i\varphi} & \epsilon_{1\downarrow}(\mathbf{k}) & 0 & J \\
J^{\ast} & 0 & \epsilon_{2\uparrow}(\mathbf{k}) & \Omega e^{i\varphi} \\
0 & J^{\ast} & \Omega e^{-i\varphi} & \epsilon_{2\downarrow}(\mathbf{k}) \\
\end{pmatrix}\,,
\label{H_tilde-matrix}
\end{equation}
where the diagonal elements  provide the SOC due to the recoil momentum  $\sqrt{2}\kappa$:
\begin{equation}
\left\{
\begin{aligned}
& \epsilon_{1\uparrow}(\mathbf{k})=[k_{x}^{2}+(k_{y}-\kappa)^{2}]/2m-\mu \\
& \epsilon_{1\downarrow}(\mathbf{k})=[(k_{x}-\kappa)^{2}+k_{y}^{2}]/2m-\mu  \\
& \epsilon_{2\uparrow}(\mathbf{k})=[(k_{x}+\kappa)^{2}+k_{y}^{2}]/2m-\mu \\
& \epsilon_{2\downarrow}(\mathbf{k})=[k_{x}^{2}+(k_{y}+\kappa)^{2}]/2m-\mu \\
\end{aligned}
\right.\,.
\end{equation}
Note that the Hamiltonian   (\ref{H_tilde-matrix}) is related  via a unitary transformation to the original Hamiltonian containing a position dependent Raman coupling  $\Omega e^{ i[\kappa(x-y)\pm i \varphi]}$ and a position-dependent interlayer tunneling $J e^{ i\kappa(x+y)}$ \cite{Su2}.

The Hamiltonian given by Eq. (\ref{H_tilde-matrix}) looks different from the one describing the SOC of the pure Rashba (or Dresselhaus) type. To establish a relation with the Rashba SOC, we set $\Omega=J$ and diagonalize the Hamiltonian (\ref{H_tilde-matrix}) for $k=0$ via the following unitary transformation:
\begin{equation}
S=\frac{1}{2}
\begin{pmatrix}
-1 & \frac{1-e^{-i\varphi}}{\sqrt{2-2\cos(\varphi)}} & \frac{e^{-i\varphi}-1}{\sqrt{2-2\cos(\varphi)}} & 1 \\
1 & \frac{1+e^{-i\varphi}}{\sqrt{2+2\cos(\varphi)}} & \frac{1+e^{-i\varphi}}{\sqrt{2+2\cos(\varphi)}} & 1 \\
-1 & \frac{e^{-i\varphi}-1}{\sqrt{2-2\cos(\varphi)}} & \frac{1-e^{-i\varphi}}{\sqrt{2-2\cos(\varphi)}} & 1 \\
1 & -\frac{1+e^{-i\varphi}}{\sqrt{2+2\cos(\varphi)}} & -\frac{1+e^{-i\varphi}}{\sqrt{2+2\cos(\varphi)}} & 1 \\
\end{pmatrix}\,.
\end{equation}
Thus one finds
\begin{equation}
\begin{split}
& H_{\mathrm{SO}}=SH_0S^{-1}=\\
& \begin{pmatrix}
\epsilon_{1+} & \alpha(k_{y}-ik_{x}) & 0 & \alpha(k_{y}+ik_{x})\\
\alpha(k_{y}+ik_{x}) & \epsilon_{1-} & \alpha(k_{y}-ik_{x}) & 0\\
0 & \alpha(k_{y}+ik_{x}) & \epsilon_{2+} & \alpha(k_{y}-ik_{x})\\
\alpha(k_{y}-ik_{x}) & 0 & \alpha(k_{y}+ik_{x}) & \epsilon_{2-}\\
\end{pmatrix} \,,
\end{split}
\end{equation}
where
\begin{equation}
\left\{
\begin{aligned}
& \epsilon_{1+}=\mathbf{k}^2/2m+E_{\mathrm{rec}}-\mu+2\Omega\sin(\varphi/2)\\
& \epsilon_{1-}=\mathbf{k}^2/2m+E_{\mathrm{rec}}-\mu+2\Omega\cos(\varphi/2)\\
& \epsilon_{2+}=\mathbf{k}^2/2m+E_{\mathrm{rec}}-\mu-2\Omega\sin(\varphi/2)\\
& \epsilon_{2-}=\mathbf{k}^2/2m+E_{\mathrm{rec}}-\mu-2\Omega\cos(\varphi/2)\\
\end{aligned}
\right.
\end{equation}
and $\alpha=\kappa/2m$.

By tuning the strength $\Omega$ and the phase difference $\varphi$ of the Raman coupling, one can reach a regime where $2\Omega\sin(\varphi/2)\gg E_{\mathrm{rec}}$ and $\sin(\varphi/2)\approx \cos(\varphi/2)$, so $\varphi \approx \pi/2$. Under these conditions the low-energy physics is described by two lowest energy branches $\epsilon_{2\pm}$ characterized by the eigen-vectors $|2\pm\rangle$. Neglecting the upper two dispersion branches $\epsilon_{1\pm}$ separated from $\epsilon_{2\pm}$ by approximately $4\Omega$, one arrives at an effective low-energy $2\times 2$ matrix Hamiltonian:
\begin{equation}
H_{\mathrm{eff}}=
\begin{pmatrix}
\mathbf{k}^2/2m-\mu'-h_{z} & \alpha(k_{y}-ik_{x})\\
\alpha(k_{y}+ik_{x}) & \mathbf{k}^2/2m-\mu'+h_{z}\\
\end{pmatrix}
\label{H_eff-AppA}
\end{equation}
where $h_{z}$ is an effective Zeeman term generated by a slight change of the phase around $\varphi=\pi/2$ \cite{Campbell}:
\begin{equation}
h_{z}=\Omega \left[\sin(\varphi/2)-\cos(\varphi/2)\right]
=\Omega \sqrt{2} \sin{\left( \delta \varphi /2\right)}\,,
\label{H_z}
\end{equation}
with $\delta \varphi =  \varphi -\pi/2$.
Here $\mu'=\mu-\Delta\mu$ is an effective chemical potential with $\Delta \mu=E_{\mathrm{rec}}-\Omega\left[\sin(\varphi/2)+\cos(\varphi/2)\right]$.

The Hamiltonian (\ref{H_eff-AppA}) can be cast in terms of the unit $2\times2$ matrix $I$ and three Pauli matrices $\sigma_{x,y,z}$:
\begin{equation}
H_{\mathrm{eff}}=
\left(\mathbf{k}^2/2m-\mu' \right)I+ \alpha\left( k_{x}\sigma_{y}+k_{y}\sigma_{x}  \right)+h_{z}\sigma_{z}\,.
\label{H_eff-1-AppA}
\end{equation}
The SOC term $k_{x}\sigma_{y}+k_{y}\sigma_{x}$ is equivalent to the Rashba SOC term $-k_{x}\sigma_{y}+k_{y}\sigma_{x}$ after interchanging the dressed states $|2+\rangle \leftrightarrow |2-\rangle$. In fact, such an interchange leads to: $\sigma_{y} \rightarrow -\sigma_{y}$ and $\sigma_{x,z}\rightarrow \sigma_{x,z}$.

Unlike in the previous study on the bilayer bosons \cite{Su2}, we do not take the phase $\varphi$ to be $\pi/2$ in the effective  Hamiltonian $H_{\mathrm{eff}}$. This allows one to include the Zeeman term playing an important role in the topological superfluidity.

\end{section}
\begin{section}{calculation of the Chern number} \label{AppendixB}

In order to calculate the Chern number, we assume that the whole region is homogenous for a sufficiently weak harmonic trap, and transform the BdG Hamiltonian to the momentum space: $H=\sum_{\mathbf{k}}\frac{1}{2}\phi_{\mathbf{k}}^{\dagger}H_{\mathrm{BdG}}(\mathbf{k})\phi_{\mathbf{k}}$, where
\begin{equation}
H_{\mathrm{BdG}}(\mathbf{k})=
\begin{pmatrix}
H_{1}(\mathbf{k}) & \mathcal{J} \\
\mathcal{J}^{\dagger} & H_{2}(\mathbf{k})\\
\end{pmatrix}\,,
\end{equation}
with
\begin{equation}
\begin{split}
H_{1}(\mathbf{k}) &=
\begin{pmatrix}
\epsilon_{1\uparrow}(\mathbf{k}) & \Omega e^{-i\varphi} & 0 & \Delta_{1}\\
\Omega e^{i\varphi} & \epsilon_{1\downarrow}(\mathbf{k}) & -\Delta_{1} & 0\\
0 & -\Delta_{1}^{\ast} & -\epsilon_{1\uparrow}(\mathbf{-k}) & -\Omega e^{-i\varphi}\\
\Delta_{1}^{\ast} & 0 & -\Omega e^{i\varphi} & -\epsilon_{1\downarrow}(\mathbf{-k})\\
\end{pmatrix}\,,\\
\end{split}
\end{equation}
and
\begin{equation}
\begin{split}
H_{2}(\mathbf{k}) &=
\begin{pmatrix}
\epsilon_{2\uparrow}(\mathbf{k}) & \Omega e^{i\varphi} & 0 & \Delta_{2}\\
\Omega e^{-i\varphi} & \epsilon_{2\downarrow}(\mathbf{k}) & -\Delta_{2} & 0\\
0 & -\Delta_{2}^{\ast} & -\epsilon_{2\uparrow}(\mathbf{-k}) & -\Omega e^{i\varphi}\\
\Delta_{2}^{\ast} & 0 & -\Omega e^{-i\varphi} & -\epsilon_{2\downarrow}(\mathbf{-k})\\
\end{pmatrix}\,.\\
\end{split}
\label{BdGequationk}
\end{equation}
Here we have chosen the Nambu basis $\phi_{\mathbf{k}}=[c_{1\uparrow,\mathbf{k}},c_{1\downarrow,\mathbf{k}},
c^{\dagger}_{1\uparrow,-\mathbf{k}},c^{\dagger}_{1\downarrow,-\mathbf{k}},c_{2\uparrow,\mathbf{k}},c_{2\downarrow,\mathbf{k}},
c^{\dagger}_{2\uparrow,-\mathbf{k}},c^{\dagger}_{2\downarrow,-\mathbf{k}}]^{T}$.
In order to determine the topological character, we then proceed to calculate the Chern number $C=1/2\pi\int d\mathbf{k}^2\Omega(\mathbf{k})$, where the $\Omega(\mathbf{k})$
is the usual Berry curvature for the momentum states \cite{XiaoD}
\begin{eqnarray}
\Omega(\mathbf{k})&=&-2\sum_{n}\sum_{m\neq
n}f_{n}\nonumber\\
&\times&\mathrm{Im}\frac{\langle\psi_{n}(\mathbf{k})|v_{k_{x}}
|\psi_{m}(\mathbf{k})\rangle\langle\psi_{m}(\mathbf{k})|v_{k_{y}}|
\psi_{n}(\mathbf{k})\rangle}{[\varepsilon_{m}(\mathbf{k})
-\varepsilon_{n}(\mathbf{k})]^{2}}\,.\nonumber\\
\label{Berrycurve}
\end{eqnarray}
Here $f_{n}=1/[e^{\varepsilon_{n}/k_{B}T}+1]$ is the Fermi-Dirac distribution function, a subscript ``$n$''($n=1,2,\cdots 8$) labels all eight particle-hole bands of the momentum-space BdG Hamiltonian, and
$\psi_{n}(\mathbf{k})$ is a wave function of eigen-energy
$\varepsilon_{n}(\mathbf{k})$, with $v_{k_{x}}$ and $v_{k_{y}}$ being
velocity operators.

In Fig.~\ref{fig4} the order parameter and the chemical potential are given by the self-consistent solution of the BdG equation (\ref{H_BdG-Eq}) at the center of the trap where the changes in the trapping potential are minimum. For the  NS and TS phases the order parameter is constant, so the momentum representation is relevant.

\end{section}

\end{document}